\documentclass[aps,pra,amsmath,amssymb,twocolumn]{revtex4}
\usepackage{amssymb}
\usepackage{color}
\usepackage{graphicx}
\usepackage{epsfig,amssymb,amsmath}
\usepackage{chngcntr}
\usepackage[normalem]{ulem}

\usepackage{color}

\newcommand{\red}{\color{red}}

\newcommand{\green}{\color{green}} 

\newcommand{\ak}[1]{{\color{red}{{#1}}}}
\newcommand{\ks}[1]{{\color{blue}{{#1}}}}

\def\>{\rangle}
\def\<{\langle}
\def\comment#1{ [{\bf Comment:} {\sf #1}]}
\def\labell#1{\label{#1}}
\def\togli#1{}

\begin{document}

\title{Quantum measurements of time} \author{Lorenzo Maccone$^1$,
  Krzysztof Sacha$^2$}\affiliation{{$1.$~Dip.~Fisica and INFN Suez.\
    Pavia, University of Pavia, via Bassi 6, I-27100 Pavia,
    Italy}\\{$2.$~Instytut Fizyki imienia Mariana Smoluchowskiego,
    Uniwersytet Jagiello\'nski, ulica Profesora Stanis\l{}awa
    \L{}ojasiewicza 11, PL-30-348 Krak\'ow, Poland}} {\begin{abstract}
    We propose a time-of-arrival operator in quantum mechanics by
    conditioning on a quantum clock. This allows us to bypass some of
    the problems of previous proposals, and to obtain a Hermitian time
    of arrival operator whose probability distribution arises from the
    Born rule and which has a clear physical interpretation. The same
    procedure can be employed to measure the ``time at which some
    event happens'' for arbitrary events (and not just specifically
    for the arrival time of a particle).
\end{abstract}}
\maketitle

Textbook quantum mechanics cannot describe measurements of time, since
time is a parameter and not a quantum observable \cite{peresbook}.
This is a clear shortcoming of the theory, since time measurements are
routinely carried out in laboratories using quantum systems that act
as clocks. Clever and creative tricks were devised to
overcome this shortcoming, e.g.~see reviews in
\cite{wernerreview,mugareview,muga2,mielnik}. However, many of these
proposals give conflicting predictions and none of them provides a
prescription that applies to generic time measurements: they all focus
on specific measurements, e.g.~the time of arrival, at a given
position, of a particle subject to a specific potential,
e.g.~\cite{rovelli,kijioski,muga,leon,galapon}. In this paper we
provide a general prescription for quantum measurements of the time at
which an arbitrary event happens (the time of arrival being a specific
instance). It entails quantizing the temporal reference frame, namely
employing a quantum system as clock
\cite{paw,aharonovt,morse,precursori,qtime}.  Then, textbook quantum
mechanics can be applied to describe time measurements through joint
quantum observables of the system under analysis and the quantum
clock.  A simple Bayes conditioning of the Born rule probability of
the joint state allows one to recover the full distribution of the
time measurement.

It is not always recognized that, in the usual formulation of quantum
mechanics, time is a conditioned quantity. The state $|\psi({t})\>$ is
the state of the system {\em conditioned} on the time being $t$ in the
Schr\"odinger picture. (Analogously, in the Heisenberg picture the
conditioning is on the observables.) This implies that the Born rule
refers to conditional probabilities: the probability that the property
$O=\sum_io_i|o_i\>\<o_i|$ has value $o_i$ is
$p(o_i|{t})=|\<\psi({t})|o_i\>|^2$, where $o_i$ and $|o_i\>$ are
eigenvalues and eigenvectors of $O$. It is a conditional probability,
{\em conditioned} on the time being $t$. Because of this, time appears
as a parameter and not as an observable in the usual formulation of
quantum mechanics \cite{peresbook}, and textbook quantum mechanics
does not directly give a quantum description of time measurements,
e.g.~the arrival time of a particle at some position
\cite{wernerreview,rovelli,leon,giannit,muga,mugareview,muga2,mielnik,werner,kijioski,holevo,dipankar,aharonov,leon1,galapon,galapon1,galapon2}.
All the previous works considered the time of arrival as a
property of the particle, and hence its corresponding observable as an
operator acting on the particle's Hilbert space (either a self-adjoint
operator \cite{rovelli,kijioski,muga} or a positive operator valued
measure (POVM) \cite{werner,giannit}). Here, instead, we consider it
as a joint property of the particle and of the clock that is used to
measure time. It is an elegant way to tackle the conditioning
described above.  Indeed, it avoids many of the technicalities of
previous proposals for time quantum observables (e.g.~the distinctions
between the interacting and the non-interacting case \cite{leon}).
Here we will consider the time operator as the one obtained from the
quantum clock.  Different systems track time in different ways and
laboratory-grade time measurements will employ the most accurate
clocks which are typically macroscopic (classical) systems. Their
energy spectrum well approximates a continuous unbounded spectrum, as
is necessary to define a good time operator. Thus the approach used in
conventional quantum mechanics of considering time as a conditioned
classical parameter is well justified in practice \cite{hartle}.
However, a fully consistent theory must possess a prescription also
for time measurements (not just as an approximation in the classical
limit), and this is what we propose here.

In this paper we use a quantum reference frame (a quantum clock) to
describe time, the Page and Wootters formalism introduced in
\cite{paw,aharonovt,morse,precursori,qtime}. This allows us to obtain
a description of the measurement of the time at which an event happens
which bypasses most problems of previous proposals.  Our proposal does
not supersede previous ones, which are well suited when considering
time as a property of the system itself, and not as a property of a
reference (the clock). However, our proposal is simple, when compared
to others: this allows us to extend it to situations beyond
time-of-arrival that other proposals cannot treat and makes it
appealing for experimental tests. While we employ a (slight) extension
of textbook quantum theory, that theory can be recovered by
conditioning the system state on what the clock
shows~\cite{paw,aharonovt,morse,precursori,qtime}.

For the sake of definiteness, we first focus on the description of the
time of arrival of a 1-dimensional particle at a position $D$ (the
position of the detector), and then show how the same mechanism can be
easily extended to the measurement of the time of occurrence of any
event. The particle at time $t$ is described by its state
$|\psi({t})\>$ and the time reference is a continuous quantum degree
of freedom described by a Hilbert space ${\cal H}_c$ with
delta-normalized basis states $|{t}\>$. We can write the global state
of particle plus reference as
\cite{paw,aharonovt,morse,precursori,qtime}
\begin{align}
  |\Psi\>=\tfrac 1{\sqrt{T}} \int_{{T}}dt\:|{t}\>|\psi({t})\>
\labell{pawst}\;,
\end{align}
which is a state in the Hilbert space ${\cal H}_c\otimes{\cal H}_s$,
with ${\cal H}_s$ the system's Hilbert space, and where the integral
is performed on the time interval $T$ (from $-\tfrac{{T}}2$ to
$\tfrac{{T}}2$), a regularization parameter\togli{
  \footnote{Non-uniform regularizations lose the ability of obtaining
    the Schr\"odinger equation \cite{qtime}, so the uniform one seems
    preferable: physically it represents all instants of time treated
    equivalently.  \togli{\ak{\sout{(time-translation
          invariance).}}}}}. One can think of $T$ as a time interval
much larger than all the other time scales: namely the physical
description of the system will be accurate for time intervals $\ll T$
and for energy intervals $\gg 2\pi\hbar/T$ (see the supplemental
material for a review of the Page and Wootters theory and for a
justification of $T$). The $\sqrt{{T}}$ term is introduced in order to
have a normalized $|\Psi\>$, given that $\<\psi({t})|\psi({t})\>=1$
for all $t$. The entanglement in \eqref{pawst} is not a result of any
clock-system dynamics (they are isolated from each other), but arises
solely from the fact that $|\psi\>$ is an eigenstate of a constraint
equation \cite{paw} (see the supplemental material).  The conventional formulation of quantum mechanics arises
from conditioning the reference to time $t$
\cite{hartle,paw,aharonovt,morse,precursori,qtime}: indeed projecting
the reference on the state $|{t}\>$ that indicates that time is $t$,
one obtains the ``state of the system given that time is $t$'':
\begin{align}
\<{t}|\Psi\>\propto |\psi({t})\>
\labell{paps}\;,
\end{align}
and the corresponding wave function $\psi(x|{t})=\<x|\psi({t})\>$ can
be obtained by projecting the system on the position eigenbasis
$\{|x\>\}$. The physical meaning of the conditioning \eqref{paps} is
that, once the clock is read out, the system conditioned on the clock
outcome is described by the state $|\psi({t})\>$, whereas all clock
outcomes are equally likely, as expected from a (uniform) quantum
clock that does not favor any particular time. We emphasize the
conditioned nature of the wavefunction by using the Bayes notation for
the conditional probability. \togli{(This approach should not be
  confused with Stueckelberg's one \cite{stu}, see
  App.~\ref{s:stuc}.)}

Since $|\Psi\>$ is normalized, it has a probabilistic interpretation
that entails a (slight) extension of the Born rule: it uses the
``history'' state $|\Psi\>$ that contains the state of the system at
all times, instead of using the state of the system $|\psi({t})\>$ at
time $t$. Indeed one can construct a time of arrival POVM as
\begin{align}\forall {t}: \Pi_{t}\equiv|{t}\>\<{t}|\otimes P_d\:;\quad
  \Pi_{na}=\openone-\int dt\:\Pi_{t}
\labell{tapovm}\;
\end{align}
where $P_d=\int_Ddx|x\>\<x|$ is the projector of the system at the
position $D$ of the detector ($D$ being the spatial interval occupied
by it). This projective POVM returns the value $t$ of the clock if the
particle is in $D$ or the value $na$ (not arrived) if it is not.  One
can easily introduce an arrival observable from it, as
\begin{align} {A}=\int dt\: {t}\:|{t}\>\<{t}|\otimes
  P_d+\openone_c\otimes\lambda\int_{x\notin D}dx\:|x\>\<x|
\labell{obs}\;,
\end{align}
where $\lambda$ is an (arbitrary) eigenvalue that signals that the
particle has not arrived (it will be dropped below by considering a
vector-valued observable) and $\openone_c$ is the identity on the
clock Hilbert space. 
The above POVM (or the observable ${A}$) does not return the
probability distribution of the time of arrival, because it also
considers the case in which the particle has not arrived. Indeed,
using it in the Born rule, one obtains the {\em joint} probability
that the particle has arrived {\em and} that its time of arrival is
$t$ (i.e. the particle is at $x\in D$ and the clock shows $t$):
\begin{align}
p({t},x\in D)=\mbox{Tr}[|\Psi\>\<\Psi|\:\Pi_{t}]=
\tfrac 1T\int_{x\in D}dx\:|\psi(x|{t})|^2
\labell{joint}\;.
\end{align}
(The case of a pointlike detector is $p=|\psi(D|{t})|^2/T$.) Then, the
time of arrival distribution is recovered from the joint probability
through the Bayes rule as
\begin{align}
  p({t}|x\in D)=\int_{x\in D}dx\:|\psi(x|{t})|^2\Big/\int_{T}
  dt\int_{x\in D}dx\:|\psi(x|{t})|^2
\labell{tdistrib},
\end{align}
where the denominator (which is the dwell time \cite{damborena}),
divided by $T$, is the unconditioned probability that the particle is
found in $D$ at any time, and $p({t}|x\in D)$ is normalized when
integrated on the interval $T$.  The dependence on the regularization
parameter $T$ disappears from the probability distribution
\eqref{tdistrib} if the time integral converges for ${T}\to\infty$,
typically if $\psi(x|{t})\to 0$ sufficiently fast in this limit. In
all other situations, the time distribution $p({t}|x\in D)$ cannot be
normalized over the whole time axis. While this might appear as a flaw
of our proposal (as it does not satisfy Kijowski's normalization axiom
\cite{kijioski}), it is actually a feature because it allows our
distribution to treat situations where Kijowski's fails, e.g.~the case
in which the particle never arrives at $D$ or the case in which the
particle is stationary at the detector: in this case $p({t}|x\in D)$
is a constant and can be normalized only on a finite interval $T$. One
can dismiss this situation as uninteresting in the classical case, but
due to the quantum superposition principle, one cannot ignore it in
the quantum case, where most reasonable wavepackets have a nonzero
probability amplitude of being stationary at the detector position:
thus, the experimental predictions of our proposal differ from the
others in this case \cite{futuro}.

Consider some other special cases: (i)~If the particle never reaches
$D$ the probability \eqref{tdistrib} is meaningless, as expected,
since the numerator and denominator are null.  (ii)~If the particle
crosses $D$ multiple times, the distribution will have multiple peaks,
as expected, corresponding to the ``crossing times''. (iii)~If the
particle is stationary at the detector position or if it reaches it
after some finite time interval but remains there forever, then
$\psi(x|{t})$ is nonzero for large $t$ and one cannot extend the time
integration to infinity.  In this case, $p({t}|x\in D)$ explicitly
depends on the interval $T$, since it is nonzero for arbitrarily large
$t$ as expected: the particle will be found at the detector at any
sufficiently large time.  (iv)~A particle performing periodic
evolution (e.g.~a harmonic oscillator) is a combination of the
previous two cases: whatever $T$ interval yields a multi-peak
distribution, but again $T$ cannot be increased to encompass all times
$t$ for which $\psi(x|{t})$ is substantially nonzero.  (v)~in the
simple case of a free nonrelativistic particle with Gaussian initial
wavepacket prepared far from the detector and negligible negative
momentum components, we obtain \cite{futuro} the same results of
\cite{kijioski,rovelli}, as expected. (vi)~If the particle is split
into two wavepackets approaching the detector from opposite
directions, our probability distribution will display interference
peaks due to the superposition principle, in contrast to other
proposals \cite{kijioski,rovelli} that do not \cite{futuro}. This
difference may be used to experimentally test our proposal.
\togli{\green (v)~If the particle is in a quantum superposition of
  multiple locations (e.g.~it has been diffracted) and will arrive in
  $D$ at a coherent superposition of different times, again the
  distribution will have multiple peaks corresponding to its
  ``collapse'' at $D$ at the corresponding times (assuming that all
  wavefunction branches arrive at the detector $D$ at some time), as
  expected: this is an immediate consequence of linearity and of the
  fact that the distribution follows from the Born rule \eqref{joint}.
  (If not all the branches arrive at the detector, the distribution
  has only the peaks relative to the arrival times of the ones that do
  arrive.)  Namely, the observable ${A}$ satisfies the superposition
  principle.{\red \bf [KS: I suggest simplifying the following text. I
    wouldn't use nomenclature of "a particle being in quantum
    superposition of multiple locations" because it is all included in
    a particle wavefunction $\psi(x|t)$.]}  }

All the above cases refer to single-shot measurements, where the time
of arrival distribution refers to the probability of outcome of a {\em
  single} measurement, which is the usual interpretation of the
Born-rule probability. Nonetheless, our mechanism can be extended to
describe also multiple (successive) measurements on the same system.
It takes into account measurement-feedback and returns the multi-time
correlations, a problem that was apparently not even ever considered
in previous literature, see \cite{qtime} and supplemental material.
Eq.~\eqref{tdistrib} is our main result.

{\em Discussion.} In previous literature, time observables are
typically defined on the Hilbert space of the system (e.g.
\cite{rovelli,holevo,kijioski,werner,galapon}). The only way the
observables of these previous proposals can give rise to a
time-of-arrival distribution through the Born rule
$p(o_i|{t})=|\<\psi({t})|o_i\>|^2$ is by postulating that the
observable is a constant of motion, such as Rovelli's evolving
constants of motion \cite{rovelliconst1,reirov}. This is a consequence
of the Born rule's conditioned nature: it contains the state at one
time only in the Schr\"odinger picture (or the observables at one time
only in the Heisenberg picture). Namely, a time of arrival operator
$\hat {t}=\int d\tau\:{\tau}\:|{\tau}\>\<{\tau}|$ with eigenstates
$|{\tau}\>$ {\em in the particle's Hilbert space} must return the same
outcome $\tau$ at any time $t$ through the Born rule:
$p(\tau|{t})=|\<\psi({t})|\tau\>|^2=\<\psi|\tau({t})\>|^2$ (in the
Schr\"odinger and Heisenberg picture respectively). This requirement
leads to awkward statements such as ``the time of arrival is $\tau$ at
time $t$'' and seems physically bizarre, since a constant of motion
should give the same outcome {\em whenever} it is measured
\cite{galapon}, but this is not the case for typical time-of-arrival
experiments, which give an outcome at a well defined time: the
measurement cannot be performed before or after the particle has
arrived. In contrast, our proposal does not suffer from this problem:
our observable is not a constant of motion.  It is a {\em joint}
observable on the system and on the clock. Its time invariance is
enforced not dynamically, but by the fact that the state $|\Psi\>$ of
system plus clock is an eigenstate of the global Hamiltonian that
defines the total energy constraint \cite{paw}.  Moreover, the
particle does not have to ``stop'' the clock nor interact with it
\cite{aharonov,unruht} as there is no clock-system interaction in the
global Hamiltonian that defines the constraint. The entanglement in
Eq.~\eqref{pawst} does not arise from a clock-system dynamics, but it
is intrinsic in the zero-energy eigenstate of the global Hamiltonian
(see supplemental material and
\cite{paw,aharonovt,morse,precursori,qtime}). Indeed, a good clock
should be {\em independent} from the system it is timing, and it is
the experimentalist that notes the correlations between particle and
clock. In other words, the conventional approach of considering the
time as a property of the system, described by an operator acting on
the system Hilbert space, is more appropriate if the system itself is
used to measure time as in many traditional approaches
(e.g.~\cite{muga2,galapon}): considering the system energy as the
generator of time translations implies that {\em that} time operator
refers to the system's evolution, but this is not what happens
typically in a lab, where experiments are timed through an external
clock. Indeed, the time operator $\int dt\:{t}\:|{t}\>\<{t}|$ in
Eq.~\eqref{obs} (where $|{t}\>$ is an eigenstate {\em in the clock
  Hilbert space}) is conjugated to the energy {\em of the clock} and
not of the system \cite{galapon,pauli}).  This is the main advantage
of our proposal, which also satisfies the desiderata for a time of
arrival operator \cite{rovelli}: it obeys the superposition principle
(trivially from linearity) and it originates from the Born rule.
Finally, a definition of time through a quantum clock can describe
real-life situations and experiments, once decoherence is accounted
for \cite{qtime}, as discussed below. Elsewhere \cite{pauli,qtime} we
have shown how the usual objections against time quantization are
overcome (see also the supplemental material).

{\em Environment and multiple clocks.} Our (idealized)
description of Eq.~\eqref{pawst} requires the system to be correlated
with a single clock and the joint system-clock state to be in a pure
state. While this is sufficient to give a fundamental prescription for
time measurements in quantum mechanics, we need to show that this is
compatible also with real world scenarios where multiple clocks are
present and where there is an environment that may interact with the
system and clocks. A less idealized description replaces
Eq.~\eqref{pawst} with 
\begin{align}
  |\Phi\>=\tfrac 1{\sqrt{T}}
  \int_{{T}}dt\:|{t}\>_{c_1}|{t}\>_{c_2}\sum_k\mu_k|\phi_k({t})\>_s|e_k({t})\>_e 
\labell{pawstenv}\;,
\end{align}
where $c_1,c_2$ indicate two different clock systems that are
synchronized (they track each other because the joint measurement of
both returns the same outcome $t$), $|\phi_k({t})\>_s$ indicates the
system state at time $t$, which may be entangled with the orthonormal
states $|e_k({t})\>_e$ of the environment with amplitude $\mu_k$.  The
time entanglement of \eqref{pawstenv} is of GHZ-type, which is the one
present in the branching states typically used in decoherence models
\cite{zurek}.  Eq.~\eqref{pawstenv} is still idealized: good clocks
should be sufficiently isolated from the environment (at least for the
time interval in which it is considered a good clock) so that their
evolution is unperturbed by it. This state contains correlations in
time even when one considers only the reduced state $\rho_{c_1s}$ of
the system and of one of the clocks, say $c_1$. In this case, the
reduced state is
\begin{eqnarray}
\rho_{c_1s}=\mbox{Tr}_{c_2e}[|\Phi\>\<\Phi|]=
\tfrac 1{{T}}  \int_{{T}}dt\:
|{t}\>_{c_1}\<{t}|\otimes\rho_s({t})
\labell{reduc}\;,
\end{eqnarray}
with $\rho_s({t})$ the reduced system state. One can obtain all
previous results using the Born rule for $\rho_{c_1s}$.
Interestingly, even though this state has lost quantum coherence in
the time correlations, retaining only classical correlations, the time
of arrival distribution can still display intereference effects
\cite{futuro}. This loss of coherence translates into an effective
superselection rule that prevents the creation and detection of
superpositions of states of different times, as expected \cite{giuli}.

At first sight, the decohered state \eqref{reduc} seems inadequate to
describe our perception of time: it describes a random time $t$,
uniformly distributed in $[-\tfrac T2,\tfrac T2]$ correlated to a
state $|\phi({t})\>_s$ \footnote{This does {\em not} mean the time
  measurement is performed at a uniform random time, since the
  measurement is a joint system-clock measurement.}. However, consider
carefully our perception of time. We perceive a single instant (the
present) and the past is contained into memory degrees of freedom,
internal to some state $|\phi({t})\>_s$: St.~Augustine's ``the past is
present memory''.  Would we be able to discriminate whether the
``present'' we perceive is a continuous succession of instants of time
(as our naive intuition suggests) or as a random sampling of instants
as described by \eqref{reduc}? No. There is no experimentally testable
way to do that.  Then such question is unscientific, and we can
conclude that \eqref{reduc} is a good description of our perception of
time even if it contains a random time. A more detailed discussion,
with a review of previous literature, is in supplemental
material.

{\em Time of arbitrary events.} Up to now we have considered the time
of arrival, which is connected to the event ``the particle is at
position $D$''. But the whole discussion can be easily extended to the
``time at which any event happens''. For example, if one considers a
spin instead of a particle on a line, one can define the ``time at
which the spin is up'', by substituting $P_d$ with the spin-up
projector $|{\uparrow}\>\<{\uparrow}|$ in Eq.~\eqref{tapovm} and by
replacing Eq.~\eqref{tdistrib} with the conditional probability that
time is $t$ given that the spin was up:
$p({t}|{\uparrow})=|\psi({\uparrow}|{t})|^2/\int_Tdt\:|\psi({\uparrow}|{t})|^2$
with $\psi({\uparrow}|{t})\equiv\<{\uparrow}|\psi({t})\>$ the
probability amplitude of having spin up at time $t$. The general case
of arbitrary events follows straightforwardly, by using a projector
$P$ that represents whatever value of a system property one wants to
consider at time $t$, namely $P$ projects onto the eigenspace relative
to some eigenvalue of a system observable. In this case the time
distribution is \begin{eqnarray}
  p({t}|P)=\mbox{Tr}[|\psi({t})\>\<\psi({t})|P]/\int_{T}
  dt\;\mbox{Tr}[|\psi({t})\>\<\psi({t})|P]
\labell{tdistrib1},\:
\end{eqnarray}
where again the dependence on the regularization parameter $T$
disappears if the integral can be taken on an interval containing all
times when the integrand is substantially different from zero and {\em
  must} be retained otherwise.  This captures the notion of ``event''
in quantum mechanics defined as ``something that happens to a quantum
system'', where ``something'' means ``a system observable property
taking some value $P$''.

{\em Expectation values and uncertainty.} The presence of $\lambda$ in
the definition \eqref{obs} of ${A}$ implies that its expectation value
$\<{A}\>=\<\Psi|{A}|\Psi\>$ is not the average
time of arrival: the observable must account also for the case in
which the particle does not arrive (or the event $P$ does not happen). 
We can partially amend by considering a 2d vector-valued observable,
where the first component of the vector contains the event time
occurrence and the second component takes care of the cases in which
the event does not occur: \togli{\red\bf [KS: I am not sure if the introduction of $\hat T$ is helpful? In App.~D $\hat T$ is called the time operator -- why? I think that we need only $\hat T_1$ (or $\hat T_{ar}$ with $\lambda=0$) and argumentation that we look for its average value provided the particle has arrived at $x\in D$ (or an event $P$ has happened). This part of the text needs improvement, in my opinion.]}
\begin{align}
\hat {T}\equiv\left(\begin{matrix}1\cr0\end{matrix}\right)\int dt\: {t}\:|{t}\>\<{t}|\otimes
  P+\left(\begin{matrix}0\cr1\end{matrix}\right)\openone_c\otimes(\openone_s-P)
\labell{that}\;.
\end{align}
\togli{\red [KS: I have impression that introducing $\hat T$ is
  redundant because what we actually need is the first part of $\hat
  T$ or the first part of $A$. My feeling is that a reader can be a
  bit confused why we are introducing an additional quantity while
  what we need is just a part of the observable $A$ which has been
  already defined. Anyway it is a minor issue and if you feel it is
  more strict to introduce $\hat T$ there is no problem.]}  The
expectation value of the first component, which we denote by $\hat
T_1$, is then proportional to the average event occurrence time
\begin{align} {t}_{ev}\equiv\alpha\<\hat{T}_1\>=\alpha\<\Psi|\int dt\:
  {t}\:|{t}\>\<{t}|\otimes P|\Psi\>
\labell{aver}\;.
\end{align}
The proportionality constant $\alpha$, arising from the Bayes
conditioning \eqref{tdistrib1}, is
$\alpha=1/\int_Tdt\:$Tr$[P|\Psi\>\<\Psi|]$.  Indeed, with this choice,
we find the correct ${t}_{ev}=\int dt\:{t}\:p({t}|P)$. It is then
clear that a null $\<{T}_1\>$ may not lead to a null ${t}_{ev}$ if the
event never happens, as in this case $\alpha=\infty$. (We can be sure
that there is a nonzero chance that the event happens if the second
component's expectation value is  $\<\hat
{T}_2\>\neq 1$.)

The presence of the constant $\alpha$ precludes the use of the
Robertson \cite{robertson} prescription to obtain a time-energy
uncertainty relation for the event occurrence time ${t}_{ev}$, since
its variance is $\Delta
{t}_{ev}^2=\alpha\<{T}_1^2\>-\alpha^2\<{T}_1\>^2$.

{\em Conclusions.} In conclusion, we have introduced a prescription
for the time measurement of when arbitrary events happen, such as the
time of arrival, by considering an observable acting on the extended
Hilbert space of system plus time reference. It satisfies the desired
properties for a time operator: it is a Hermitian operator (on the
extended space), its probability distribution arises from the Born
rule, it satisfies the superposition principle and it has the correct
physical interpretation arising from a mathematical description of
what happens in an actual experiment\togli{ \footnote{At a recent foundations
  conference, an attendee asked an experimentalist (P. Kolenderski)
  ``how could you measure the photon's time of arrival? There's no
  Hermitian operator to describe such observable!'' Kolenderski simply
  replied: ``I looked at my oscilloscope''. Our proposal provides
  exactly an observable for this kind of experimental situations: an
  experimentalist correlating a system (the photon) with a clock (the
  oscilloscope).}}.

\vskip 1\baselineskip LM acknowledges funding from Unipv, ``Blue sky''
BSR1718573, the FQXi grant RFP-1513 and the Attract grant 777222.  KS
acknowledges support of the National Science Centre, Poland via
Project No. 2018/31/B/ST2/00349. We thank the University Centre
Obergurgl, Austria.

\newpage
\setcounter{equation}{0}
\counterwithout{equation}{section}


\renewcommand{\theequation}{S\arabic{equation}}
\section*{\large\sc Supplemental Material to ``Quantum measurements of time''}

\section{Page and Wootters theory and the finite time window $T$
  approximation\label{s:paw}}
Here we recall the Page and Wootters mechanism
\cite{paw,aharonovt,morse,precursori,qtime,hartle} and analyze the
approximation introduced by considering a finite time window $T$ in
Eq.~\eqref{pawst}, to avoid problems connected to regularizations (see
\cite{qtime,pauli}).

The reader that is familiar with the Page and Wootters mechanism and
constrained quantum mechanics can ignore this section. In
Sec.~\ref{s:philo} we comment on the fundamental implications of
this formalism.

While the introduction of the regularization parameter $T$ may seem
restrictive, it is a very good approximation if one considers a
sufficiently large time interval $T$. Moreover, as discussed in the
paper, it permits to retain a probabilistic interpretation also to the
cases where a normalization over the entire time axis is impossible,
e.g.~the time of arrival measurement of a particle that is stationary
at the detector.

The Page and Wootters mechanism arises by requiring that the state
$|\Psi\>$ of Eq.~\eqref{pawst} is an eigenstate of the total energy
given by the system Hamiltonian $H_s$ plus the clock Hamiltonian
(which coincides with its momentum $\Omega$):
\begin{align}
(\hbar \Omega\otimes\openone_s+\openone_c\otimes H_s)|\Psi\>=0
\labell{paweig}\;,
\end{align}
where $\openone_c$ and $\openone_s$ are the identity operators for the
clock and the system respectively.  The physical interpretation of the
constraint equation is the following. The physical states are the
eigenstates of the constraint \eqref{paweig}. For these states the
Hamiltonian of the system $H_s$ is equal to the generator of time
translations $\hbar\Omega$ (the momentum operator) of the clock:
$H_s|\psi\>=-\hbar\Omega|\psi\>$. In other words, the dynamics of the
conventional formulation of quantum mechanics is replaced by the
constraint equation.

Note that in the total Hamiltonian there is no coupling term between
clock and system: a good clock is isolated from the rest
\cite{hartle,paw}. The constraint equation can be taken as evidence of
a gauge invariance: since the physical states $|\Psi\>$ are
eigenstates of the global Hamiltonian, they are invariant for shifts
of the conjugate quantity, which can be interpreted as the
``coordinate time'' in general relativity \cite{rovelliconst1,hartle},
which is just a gauge because of the diffeomorphism invariance of that
theory. The physical time is then attached to some {\em internal}
degree of freedom, the ``clock time'', which is conjugate to the clock
Hamiltonian $\hbar\Omega$. This idea originates from Dirac's analysis
of quantum constrained systems, e.g.~see \cite{morse}. In general
relativity it is not possible in general to have a constraint of the
form {$\hbar\Omega\otimes\openone_s+\openone_c\otimes H_s$.} When this
is possible, then the constraint gives rise to the time-dependent
Schr\"odinger equation \cite{morse,paw} (we consider only
non-relativistic quantum mechanics in this paper).  In fact, in the
``position'' representation, the momentum satisfies
$\<{t}|\Omega=i\tfrac\partial{\partial {t}}\<{t}|$ and
$\<{t}|\Psi\>=|\psi({t})\>$, so Eq.~\eqref{paweig} in the clock position
representation becomes \begin{align}
  {i\hbar\tfrac\partial{\partial {t}}}|\psi({t})\>-H_s|\psi({t})\>=0
\labell{sch}\;,
\end{align}
the time-dependent Schr\"odinger equation. 

We emphasize that it is Eq.~\eqref{sch} that should be considered as
the dynamical equation that describes the time evolution of the state
with respect to (internal) time $t$ marked by the clock. Instead,
Eq.~\eqref{paweig} is only a constraint equation that tells us that
the total state $|\Psi\>$ is an eigenstate of the total Hamiltonian,
namely it does {\em not} evolve with respect to any (external) time:
all evolution is contained {\em inside} $|\Psi\>$ as an {\em internal}
degree of freedom. In particular, one must not interpret $|\Psi\>$ as
a result of some entangling dynamics: $|\Psi\>$ is a state which is
stationary with respect to all external degrees of freedom.

The con\-straint also gives rise to the time-independent Schr\"odinger
equation \cite{qtime} in the ``momentum'' representation, where
$\<\omega|\Omega=\omega\<\omega|$ and
$\<\omega|\Psi\>=|\tilde\psi(\omega)\>$ with $|\tilde\psi(\omega)\>$
proportional to the system energy eigenstate of energy $-\hbar\omega$
or $|\tilde\psi(\omega)\>=0$ if the system does not possess energy
$-\hbar\omega$ in its energy spectrum, namely Eq.~\eqref{paweig} in
the momentum representation for the clock becomes \begin{align}
  \hbar\omega|\tilde\psi(\omega)\>+H_s|\tilde\psi(\omega)\>=0
\labell{scht}\;,
\end{align}
the time-independent Schr\"odinger equation, valid only if the
Hamiltonian $H_s$ has $-\hbar\omega$ in its spectrum.  We emphasize
that the energy $\hbar\Omega$ conjugate to the time observable is the
clock energy and not the system energy (whose spectrum may be
completely arbitrary) \cite{pauli}.

The above properties hold in the ideal (idealized) case of unbounded
continuous time and energy spectrum of the clock. In this paper we
considered a clock with a continuous but bounded time spectrum
${t}\in[-{T}/2,{T}/2]$, where the eigenvalue $t$ is connected to an
eigenvector $|{t}\>$. Then its conjugate momentum (coinciding with the
clock energy \cite{paw}) has discrete spectrum  {$n2\pi/T$ corresponding to eigenstates:}
\begin{align}
|n\>=\frac1{\sqrt{{T}}}\int_Tdt\:e^{itn\tfrac{2\pi}{T}}|{t}\>,\
|{t}\>=\frac1{\sqrt{{T}}}\sum_{n=-\infty}^\infty e^{-itn\tfrac{2\pi}{T}}|n\>
\labell{nn}\;,
\end{align}
where we used $\sum_{n=-\infty}^\infty
e^{itn2\pi/{T}}=\sum_{m=-\infty}^{\infty}\delta({t}-mT)$ with $\delta$
the Dirac delta.  \togli{\comment{Check the dimensions! It's possible that
  we may need to shift the $\sqrt{{T}}$ term around to get the correct
  dimensions!}  }With these definitions, the state \eqref{pawst} in the Letter is
\begin{align}
|\Psi\>=\frac1{\sqrt{{{T}}}}\int_Tdt\:|{t}\>|\psi({t})\>=
\frac1{\sqrt{{{T}}}}\sum_{n=-\infty}^\infty|n\>|\tilde\psi_n\>
\labell{paswt2}\;,\\\mbox{with
}|\tilde\psi_n\>\equiv\tfrac1{\sqrt{{T}}}\int_Tdt\:
e^{itn\tfrac{2\pi}{T}}|\psi({t})\>.
\end{align}
The state of the system in terms of the system's energy eigenstates
$|e_m\>$ (connected to the eigenvalues $\hbar\omega_m$) is
\begin{eqnarray} &&|\psi({t})\>=\sum_m\psi_me^{-i\omega_mt}|e_m\>
\labell{eigen}\;,\\&&\Rightarrow
|\tilde\psi_n\>=\tfrac1{\sqrt{{T}}}\sum_m\psi_m|e_m\>
\int_Tdt\:e^{-it(\omega_m+n2\pi/{T})}\nonumber
\labell{ei}\;.
\end{eqnarray}
It follows that $|\tilde\psi_n\>$ is proportional to a system energy
eigenstate $|e_m\>$ only if the clock momentum $\Omega=\tfrac{2\pi}{T}
\sum_n n|n\>\<n|$ has a spectrum $\tfrac{2\pi}{T}n$ sufficiently dense
so that there exist an integer $n$ such that $\omega_m=n2\pi/T$ which
can be attained asymptotically for large $T$.  In this case the
integral in \eqref{eigen} is proportional to a Dirac delta, and the
constraint equation \eqref{paweig} gives rise to the time-independent
Schr\"odinger equation for the system by multiplying both members to
the left by the clock energy eigenstate $\<n|$:
\begin{align}
H_s\<n|\Psi\>=-\hbar\tfrac{2\pi}{T}n\<n|\Psi\>\:\Leftrightarrow\:
H_s|\tilde\psi_n\>=\hbar\omega_m|\tilde\psi_n\>
\labell{schindep}\;,
\end{align}
with $\omega_m=-{2\pi}n/T$ for some $n$ for each of the system
spectrum eigenvalues $\omega_m$ (and choosing $|\tilde\psi_n\>=0$ a
null vector if $-{2\pi}n/T$ is not a system eigenvalue).

\section{Measurements and preparations}
We now review how measurements are treated in constrained quantum
mechanics. In particular we show how the formalism yields the correct
statistics of the measurement outcomes that one expects from the Born
rule. This treatment explains why one should not necessarily be
confined to observables that are constants of motion \cite{rovelli}.
The material that follows is a simplified review of \cite{qtime}.

Suppose that we measure the observable $A=\sum_aa|a\>\<a|$ at time
${t}_a$, where $a$ are the eigenvalues and $|a\>$ the eigenvectors. To
obtain the measurement statistics, we track the interaction between
the system and the measurement apparatus, using von Neumann's
pre-measurement prescription \cite{vonneuman} and calculate the Born
rule using projectors on the degrees of freedom (the memories) that
encode the measurement outcomes. An impulsive measurement of $A$ that
takes place exactly at time ${t}_a$ is described by the von Neumann
pre-measurement unitary evolution \cite{vonneuman}
\begin{eqnarray}
|\psi({t}_a)\>{|r\>_{m}}\to\sum_a\psi_a|a\>{|a\>_m,} \mbox{ with }
\psi_a\equiv\<a|\psi({t}_a)\>
\labell{imp},\qquad
\end{eqnarray}
where $|\psi({t}_a)\>$ and {$|r\>_m$} are the initial states of 
system and memory. The
history state that describes this evolution is \cite{qtime}
\begin{eqnarray}
|\Psi\>=
\!\!\frac 1{\sqrt{{T}}}
\Big[\int_{-T}^{{t}_a}\!\!dt|{t}\>|\psi({t})\>|r\>_m\nonumber\\+
\int_{{t}_a}^{T}dt|{t}\>\sum_a\psi_a{U(t-t_a)}|a\>|a\>_m\Big]
\labell{hist}
,\labell{psimeas}
\end{eqnarray}
where $U$ is the time evolution operator for the system only (the
interaction with the memory is considered explicitly), $|\psi({t})\>$
the state prior to the measurement, and $m$ is the memory degree of
freedom where the measurement outcome is stored. The integrals in
\eqref{hist} represent the time evolution before the measurement and
after the measurement respectively. Note that ${t}_a$ is a value
referred to the internal clock even if it seems an external parameter.
The Born rule for the projective POVM that projects the memories onto
the state $|a\>_m$ gives a probability for the outcomes
\begin{align}
p(a)=\mbox{Tr}[(|a\>_m\<a|\otimes\openone)|\Psi\>\<\Psi|]
\labell{born}\;,
\end{align}
where $\openone$ is the identity that acts on the Hilbert space of the
rest (excluding the memory), and we assume that $|r\>_m$ is orthogonal
to all the $|a\>_m$. Then one immediately recovers the expected Born rule
probability \begin{align}
p(a|t)=|\psi_a|^2=|\<a|\psi({t})\>|^2
\labell{brul}\;
\end{align}
if ${t}>{t}_a$ or $p(a|{t})=0$ otherwise. Refer to \cite{qtime} for
more general situations, such as POVM measurements.

Before closing this section, we briefly comment on how this formalism
can describe the preparation of a state at a given time. A
state preparation is just a measurement, in which one specific outcome
is post-selected \cite{peresbook}. Then a Page-Wootters state that
describes the preparation of a system in the state $|\psi_0\>$ at time
${t}=0$ is the following
\begin{eqnarray}
&&  |\Psi''\>=\frac1{\sqrt{{T}}} \left[\int_{-{T}/2}^0
  dt\:|{t}\>|i\>|r\>_m\right.\labell{sada}\\\nonumber&&\left.+ \int^{{T}/2}_0
  dt\:|{t}\>\left({U(t)}|\psi_0\>|0\>_m+U_d\sum_{j\neq0}|j\>
|j\>_m\right)\right]
\;,
\end{eqnarray}
where $|i\>|r\>_m$ are the states of the system and memory before the
preparation, at time ${t}=0$ a measurement that uses the von Neumann
prescription \eqref{imp} happened, and the $m$ degree of freedom is
where the measurement outcome is stored. The measurement is chosen in
such a way that the outcome $j=0$ corresponds to the state $|\psi_0\>$
that needs to be prepared. The experimenter will discard (post-select)
all cases when the memory does not contain the outcome $j=0$ by
running a ``discard'' transformation $U_d$ (a conditional unitary,
conditioned on the measurement outcome $m$) and keep only the state
$|\psi_0\>$ corresponding to $j=0$. For example, this is the way that
a spin-$1/2$ ``up'' state can be prepared: the experimenter sends the
spin through a Stern-Gerlach apparatus and discards (using some
physical transformation $U_d$ such as a beam-stop) all spins that end
in the lower arm, post-selecting on the spins that end up in the
higher arm. These are thus prepared in the spin up state.

\section{Multiple measurements at different times\label{s:multiple}}
In this section we extend the discussion of the previous section to
the case in which multiple measurements are performed at different
times and show how the formalism yields the correct statistics of the
measurement outcomes that one expects from the Born rule. This
treatment explains why one should not necessarily be confined to
observables that are constants of motion \cite{rovelli}.  Also the
material in this section is a simplified review of \cite{qtime}.

We first remind how correlations among successive measurements at
fixed times can be obtained and then we will extend the analysis to
the case in which the times themselves are measured.

Suppose that we now measure the observable $A=\sum_aa|a\>\<a|$ at time
${t}_a$ and the observable $B=\sum_bb|b\>\<b|$ at time ${t}_b$, where
$a$ and $b$ are the respective eigenvalues and $|a\>$ and $|b\>$ the
eigenvectors. Using twice a von Neumann pre-measurement unitary of the
type \eqref{imp} for the measurement of $A$ and of $B$, we can
construct a history state that describes both evolutions \cite{qtime}
\begin{widetext} 
\begin{align}
|\Psi\>=
\!\!\int_{-\infty}^{{t}_a}\!\!dt|{t}\>|\psi({t})\>|r\>_1|r\>_2+
\int_{{t}_a}^{{t}_b}\!\!dt|{t}\>\sum_a\psi_aU({t}-{t}_a)|a\>|a\>_1|r\>_2
+\int^{\infty}_{{t}_b}\!\!dt|{t}\>\sum_{ab}\psi_aU({t}-{t}_b)|b\>\<b|
U({t}_b-{t}_a)|a\>|a\>_1|b\>_2
,\labell{hist1}\;
\end{align}
\end{widetext}
where $U$ is again the time evolution operator for the system only,
$|\psi({t})\>$ the state prior to the measurements, and $|\>_1|\>_2$
are the memory degrees of freedom where the two measurement outcomes
are stored. The three integrals in \eqref{hist1} represent the time
evolution before the first measurement, between measurements, and
after the second measurement respectively. Now the Born rule for the
projective POVM that projects the memories onto the state
$|a\>_1|b\>_2$ gives a joint probability for the outcomes given by
\begin{align}
p(a,b)=\mbox{Tr}[(|a\>_1\<a|\otimes|b\>_2\<b|\otimes\openone)|\Psi\>\<\Psi|]
\labell{born1}\;,
\end{align}
which extends \eqref{born}. Again one immediately recovers the
correct probabilities and correlations for a two time measurement from
Eq.~\eqref{hist1}: the joint probability is
$p(a,b)=|\psi_a\<b|U({t}_b-{t}_a)|a\>|^2$
and the conditional probability 
\begin{align}
p(b|a)=|\<b|U({t}_b-{t}_a)|a\>|^2,
\labell{cop1}\;
\end{align}
as expected. (An experimental realization of the above state was
performed in \cite{conttime,esperimentotorino}.)

A strong objection against the Page and Wootters formalism is that
measurements at different times seem to give the wrong multi-time
correlation functions \cite{kuchar}. The above discussion shows how
one can overcome this objection \cite{qtime}.

However, the most famous objection against a quantization of time is
the one due to Pauli: a straightforward application of the Stone-von
Neumann theorem implies that if a time operator $\hat T$ is
canonically conjugate to the system Hamiltonian $H_s$, as $[\hat
T,H_s]=i\hbar$, then the Hamiltonian must have the same (unbounded
continuous) spectrum as $\hat T$, which is typically not true. The
Page and Wootters construction is immune to this criticism, since the
time operator is an operator of the clock, whereas the system
Hamiltonian $H_s$ is an operator of the system. So, in our case $[\hat
T,\hbar\Omega]=i\hbar$ (and indeed the Hamiltonian of the clock, its
momentum, has an unbounded continuous spectrum), while $[\hat
T,H_s]=0$ since these operators act on different Hilbert spaces. Then,
there are no demands on the spectrum of the Hamiltonian $H_s$ from the
Stone-von Neumann theorem. More details on this can be found in
\cite{pauli}.

\section{Time correlations}
In this section we employ the results reviewed in the previous section
to describe what happens when one performs successive time
measurements. Apparently this problem was not even considered in
previous literature, where only single time measurements were
analyzed.

\togli{\ks{\comment{KS: I suggest to restrict to the impulse case only. In (45) the impulse case is considered and we don't need to make the present section more complicated --- of course we can mention that the analysis can be extended to the measurements that last finite times.}}}
While the history state \eqref{hist1} describes two impulsive
measurements at fixed times ${t}_a$ and ${t}_b$, the above analysis
can be extended to the case in which the measurements take a finite
amount of time and the times at which the interactions happen are
themselves copied into a further memory (a ``timestamp memory'') by
another interaction.  Namely, consider the situation where the memory
degrees of freedom store both the outcome and the time at which the
measurement is performed, by replacing \eqref{imp} with a general
(non-impulsive) interaction that starts from a factorized state
$|\psi({t})\>|r\>_1|r\>_{t1}$ for ${t}\to-\infty$ and ends with a
joint entangled state $|\theta({t})\>$ of system and memories for
${t}\to\infty$, where the entanglement in $|\theta\>$ refers to the
correlation both of the measured observable as in Eq.~\eqref{imp} {\em
  and} of the time variable, stored in the memory $t1$.  In the
impulsive limit of \eqref{imp} the unitary evolution
$|\psi({t})\>|r\>_1|r\>_{t1}\to|\theta({t})\>$ takes the form
$|\psi({t})\>|r\>_1|r\>_{t1}\to\sum_a\psi_a|a\>|a\>_1|{t}_a\>_{t1}$,
where the time-memory-register $t1$ remains factorized from the rest.
In the general case it will instead lead to an entangled history state
of the type $\int dt|{t}\>|\theta({t})\>$ that describes correlations
that build up among system and memories. \togli{WARNING!  I'm not
  sure this can be really done! Namely, I'm not sure how one can model
  the correlations that build up with the timestamp memories. I think
  they should build up exactly as when one makes an interaction with a
  time-evolving system, so I'm optimistic, but I'd like to see it done
  at least in a simple case. We should at least work it out for a
  simple example: e.g.~von Neumann's example \cite{vonneuman} pg 443
  where he builds up a measurement by considering an interaction of
  the type $qP$ where $q$ and $P$ are the position and momentum of two
  (large) particles... We would have to build TWO unitary
  interactions: one that correlates the system with the memory (that's
  what von Neumann does) and an identical one that correlates the
  clock with the memory, but is active only when the first is.}

If one wants to describe a single measurement, the degree of freedom
$t1$ that records the time is not important as one can directly look
at the clock (as we have done in the main text). It becomes crucial
whenever one wants to describe multiple measurements. In this case,
the correlations arise by comparing the measurements {\em and} the
relative timestamps. Indeed, to describe two measurements, one can
extend the above description to include the memories $1$ and $2$ where
the two outcomes are stored and the memories $t1$ and $t2$ where the
measurement times are stored, obtaining 
\begin{align}
|\Psi'\>=\tfrac1{\sqrt{{T}}}\int dt|{t}\>|\theta'({t})\>
\labell{histtt}\;,
\end{align}
where now $|\theta'({t})\>$ is a joint state of the system, and of the
four memories that store the two measurement outcomes and the two
timestamps. For {example, for} a two-particle time-of-arrival, the POVM that tells us
that particle $a$ has arrived at position $d_a$ at time ${t}_a$ and
particle $b$ at position $d_b$ at time ${t}_b$ is 
\begin{align}
\Pi_{{t}_a,{t}_b}=|{t}_a\>_{t1}\<{t}_a|\otimes P_{d_a}\otimes
|{t}_b\>_{t2}\<{t}_b|\otimes P_{d_b},\:\\
\Pi_{na}=\openone-\int
dt_adt_b\Pi_{{t}_a,{t}_b} 
\labell{tapovm1}\;,
\end{align}
which generalizes the Eq.~(3) of our paper and where the
projectors act on the time memories $t1$ and $t2$ and on the two
particles' Hilbert spaces $d_a$ and $d_b$. Note the necessity of the
timestamp memories $t1$ and $t2$ in order to define this POVM. The
POVM element $na$ reveals that at least one of the two particles has
not arrived. The joint probability that particle $a$ was found at
$d_a$ at time ${t}_a$ {\em and} particle $b$ at $d_b$ at time ${t}_b$
is obtained through the Born rule as
$p({t}_a,{t}_b,d_a,d_b)=$Tr$[|\Psi'\>\<\Psi'|\Pi_{{t}_a,{t}_b}]$.
Hence, through the Bayes rule, the probability of detecting the two
particles at time ${t}_a$ and ${t}_b$ is
\begin{align}
p({t}_a,{t}_b|d_a,d_b)=\frac{\mbox{Tr}[|\Psi'\>\<\Psi'|\Pi_{{t}_a,{t}_b}]}
{\int_T dt_adt_b\mbox{Tr}[|\Psi'\>\<\Psi'|\Pi_{{t}_a,{t}_b}]}
\labell{con}\;,
\end{align}
which generalizes Eq.~(6) of our paper.

\togli{\comment{Aggiungere il caso di multiple clocks compreso quello di
  \cite{vittorio}: {\em forse:} In a typical experiment, the clock
  $|{t}\>$ in \eqref{tapovm} that ``purifies'' the system state as in
  Eq.~\eqref{pawst} is inaccessible.  This is a not a problem, since
  the Page and Wootters construction can accommodate multiple clocks
  that track one another \cite{vittorio} and the experimentalist will
  use one of them.  Namely, the ``master'' clock system in the state
  \eqref{pawst} can be thought a collection of clocks ($a$, $b$, etc.)
  that agree with one another as $|{t}\>=|{t}\>_a|{t}\>_b\cdots$ so
  that $|\Psi\>\propto\int dt |{t}\>_a|{t}\>_b\cdots|\psi({t})\>$, and
  the experimentalist has access to only one. This does not
  substantially alter the above discussion, but the presence of
  multiple clocks implies a decoherence in the time basis analogous to
  the one obtained from quantum Darwinism \cite{zurek}.  This is
  exactly what happens in real-world scenarios whenever a
  ``classical'' clock is employed to gauge time (but the quantum
  nature of the clock may be retained in controlled scenarios
  \cite{conttime,esperimentotorino}).  }}

\section{Fundamental questions\label{s:philo}}

In this section we detail the foundational motivations that lead us to
employ the constraint formalism of Page and Wootters, and give an
interpretation of the results and of the quantum states we introduced
in the previous sections. Some of the considerations discussed here
have appeared in the literature many times previously,
e.g.~\cite{paw,aharonovt,precursori,qtime,rovelliconst1} and
elsewhere.

In textbook quantum mechanics, time is treated as a (classical)
parameter. This parameter is typically interpreted as the time shown
by a (classical) clock that is external to the system. This clock is
used to determine the time of the events described by the
Schr\"odinger equation and of the measurements described by the Born
rule. So, textbook quantum mechanics requires an external classical
world both to describe the measurement apparatus and the clocks that
establish the timings. This is perfectly consistent as long as one
does not want to give a quantum description of time measurements, so
we certainly do not advocate that textbook quantum mechanics should be
abandoned.

The fact that time is determined externally to quantum mechanics is
reflected in the fact that states and Born-rule probabilities in
textbook quantum mechanics are conditioned quantities, as discussed in
the main text: $|\psi({t})\>$ is the state of the system {\em given
  that} the time is $t$, and $p(a|{t})=|\<a|\psi({t})\>|^2$ is the
probability that the outcome of the observable $A=\sum_aa|a\>\<a|$
gives result $a$ {\em given that} the time is $t$.

Now, we want to go beyond textbook quantum mechanics and provide a
quantum description of time measurements. Clearly this
cannot be done as long as the conditioning happens on a classical
clock system. A clever idea
\cite{paw,aharonovt,morse,precursori,qtime} is to use a quantum system
as a clock. Then, one can retain most of textbook quantum mechanics
with minimal changes: (i)~one cannot use the Schr\"odinger equation
where the time variable is a classical parameter, but that equation
can be replaced by the constraint equation \eqref{paweig} which is
equivalent to the Schr\"odinger one when written in the position
representation, as shown above. (ii)~the Born rule should be applied
not to the time-conditioned state $|\psi({t})\>$, but to the
Page-Wootters history state $|\Psi\>$ that, as an eigenstate of the
constraint equation, contains the full dynamics (i.e.~the solutions of
the Schr\"odinger equation at all times).

The rest of quantum mechanics can be retained without changes, since
textbook quantum mechanics can be obtained by conditioning the system
state and the system measurement outcomes on the clock measurement
outcomes, as shown in the previous section and in \cite{qtime}.

One may object that there is no ``flow'' of time in a state such as
\eqref{psimeas}, since time is an internal degree of freedom, and it
describes a sort of ``frozen time formalism'' \cite{rovelliconst1}.
This objection can be bypassed on many levels. 

(i)~At a philosophical level, it has been proposed multiple times and
by many that time does not flow. Events flow {\em in} time (a river
flows with respect to time), but it is meaningless to assume that time
{\em itself} flows. If one were to claim that time flows in itself,
that would lead to apparently meaningless claims such as ``time flows
at a `speed' of one second per second''.  Some philosophers
(e.g.~Barbour, McTaggart)\togli{These are certainly not the only
  ones! We should probably add some references here} push this
argument to its ultimate consequence, and claim that time does not
exist. Leibnitz apparently even changed the spelling of his last name
to Leibniz to protest against the existence of time as an absolute
entity that flows irrespective of any event.  (Newton famously took
the opposite view.) We will not enter into the philosophical debate
here, but just point out that our formalism embodies the relational
view of Leibniz: time in itself does not flow, but it can be seen as a
relation between different events.  This is evident for example in the
state \eqref{psimeas}, where the temporal degree of freedom is
internal and its change can only be ascertained by correlation
measurements between that degree of freedom and other internal degrees
of freedom. A measurement of the time degree by itself would only
return a random outcome without intrinsic meaning.

(ii)~At a physical level, the possibility of having a quantum
formalism that considers time as an internal degree of freedom is very
appealing for different reasons. For example, approaches based on
constraint equations a la Dirac have been very fruitful in quantum
cosmology \cite{hartle} and in canonical approaches to the
quantization of gravity \cite{rovelliconst1,wdw}. In non-relativistic
quantum mechanics, the constraint equation can be more simply
interpreted as a global conservation law (the conservation of total
energy, in our case) which, through Noether's theorem, is connected to
a global time translational invariance: only the internal time is
meaningful, whereas the ``external'' time (the variable conjugate to
the constraint) is unphysical.

(iii)~At a conceptual level, we need to show that our formalism is
consistent with our experiences and perceptions. If one measures the
time degree of freedom in a Page-Wootters state such as
\eqref{psimeas}, one obtains a random outcome $t$ uniformly
distributed between $-\infty$ and $+\infty$ (or in the interval
$[-{T}/2,{T}/2]$ in the regularized version). This seems to blatantly
contradict our perception of time measurements, where we
perceive time as a continuous stream of connected perceptions.
However, a more careful reflection shows that this perception is not
at all in contradiction with the Page-Wootters state. What we {\em
  really} perceive is a single instant of time (the present) and we
are aware of the past only through our memories: Sidney Coleman's
``the past is present memory''. Whether or not different time
measurements that lead to our perception of time happen in a
continuous succession (as our naive perception of time suggests) is a
completely untestable statement: there is no way it can be falsified
experimentally. Hence it is not a scientific statement. This is just
the fallacy of the ``flow'' of time itself in another
guise.

Thus we can confidently claim that the Page-Wootters formalism is
foundationally appealing: it encodes Leibniz's
    relationalism; 
it has been fruitfully employed in quantum mechanics of constrained
systems and requires the minimal changes to textbook quantum mechanics
required by a quantization of the temporal degree of freedom; it is
compatible with our own perception of time.


\begin{references}
\bibitem{peresbook}A. Peres, {\em Quantum Theory: Concepts and
    Methods}, (Kluwer ac. publ., Dordrecht, 1993).
\bibitem{wernerreview}A.  Ruschhaupt, R.F. Werner, Ch. 14 Quantum
  Mechanics of Time in {\em The Message of Quantum Science, Lecture
    Notes in Physics 899} (Springer, 2015), P.  Blanchard, J.
  Fr\"ohlich (eds.).
\bibitem{mugareview}J. G. Muga, R. Sala, J. P. Palao, The time of
  arrival concept in quantum mechanics, Superlattices and
  Microstructures, {\bf 23}, 833 (1998), quant-ph/9801043.
\bibitem{muga2}J. G. Muga, R. Sala Mayato, I. Egusquiza, (Eds.)  {\em
    Time in Quantum Mechanics} (Springer Lecture Notes in Physics,
  2008).
\bibitem{mielnik}B. Mielnik, The screen problem, Found. Phys. {\bf
    24}, 1113 (1994).
\bibitem{rovelli} N. Grot, C. Rovelli, R.S. Tate, Time of arrival in
  quantum mechanics, Phys. Rev. A {\bf 54}, 4676 (1996).
\bibitem{kijioski}J. Kijowski, On the time operator in quantum
  mechanics and the Heisenberg uncertainty relation for energy and
  time, Rep. Math. Phys. {\bf 6}, 361 (1974); J. Kijowski, Comment on
  ``Arrival time in quantum mechanics'' and ``Time of arrival in
  quantum mechanics'', Phys. Rev. A {\bf 59}, 897 (1999).
\bibitem{muga}V. Delgado and J. G. Muga, Arrival time in quantum
  mechanics, Phys. Rev. A {\bf 56}, 3425 (1997).
\bibitem{leon}J. Le\'on, J. Julve, P. Pitanga, F.J. de Urr\'ies, Time
  of arrival in the presence of interactions. Phys. Rev. A {\bf 61},
  062101 (2000).  
\bibitem{galapon}E. A. Galapon, Theory of quantum arrival and spatial
  wave function collapse on the appearance of particle, Proc. R. Soc.
  A {\bf 465}, 71 (2009).
\bibitem{paw}D.N. Page and W.K. Wootters, Evolution without Evolution:
  Dynamics described by stationary observers, Phys. Rev. D {\bf 27},
  2885 (1983).
\bibitem{aharonovt}Y. Aharonov and T. Kaufherr, Quantum frames of
  reference, Phys. Rev. D {\bf 30}, 368 (1984).
\bibitem{morse}P. McCord Morse, H. Feshbach, {\em Methods of
    Theoretical Physics, Part I} (McGraw-Hill, 1953), Chap. 2.6.
\bibitem{precursori} N. Mott, Proc. Roy. Soc. A {\bf 126}, 79 (1929);
  T. Banks, Nucl. Phys. B {\bf 249}, 332 (1985); R.  Brout, Found.
  Phys. {\bf 17}, 603 (1987); R. Brout, G.  Horwitz, D.  Weil, Phys.
  Lett. B {\bf 192}, 318 (1987); R.  Brout, Z. Phys. B {\bf 68}, 339
  (1987); F. Englert, Phys. Lett. B {\bf 228}, 111 (1989); J. S.
  Briggs and J. M. Rost, Found. Phys. {\bf 31}, 693 (2001).
\bibitem{qtime}V.Giovannetti, S.Lloyd, L.Maccone, Quantum time, Phys.
  Rev. D, {\bf 92}, 045033 (2015).
\bibitem{giannit} R. Giannitrapani, Positive-Operator-Valued Time
  Observable in Quantum Mechanics, Int. J. Theor. Phys. {\bf 36}, 1575
  (1997).
\bibitem{werner}R. Werner, Screen observables in relativistic and
  nonrelativistic quantum mechanics, J.  Math. Phys. {\bf 27}, 793
  (1986).
\bibitem{holevo}A. Holevo, {\em Probabilistic and Statistical Aspects
    of Quantum Theory}, (Scuola Normale Superiore Pisa, 2011).
\bibitem{dipankar} D. Home, A. K. Pan, A. Banerjee, Quantitative
  probing of the quantum-classical transition for the arrival time
  distribution, J. Phys. A: Math. Theor. {\bf 42}, 165302 (2009).
\bibitem{aharonov}Y. Aharonov, J. Oppenheim, S. Popescu, B. Reznik,
  and W. G. Unruh, Measurement of time of arrival in quantum
  mechanics, Phys. Rev. A {\bf 57}, 4130 (1998).
\bibitem{leon1}J. Le\'on, Time-of-arrival formalism for the
  relativistic particle, J. Phys. A: Math. Gen. {\bf 30} 4791 (1997).
\bibitem{galapon1} E.A. Galapon, R.F. Caballar, R.T. Bahague Jr,
  Confined Quantum Time of Arrivals, Phys. Rev. Lett. {\bf 93}, 180406
  (2004).
\bibitem{galapon2}E. Galapon, Pauli's theorem and quantum canonical
  pairs: the consistency of a bounded, self-adjoint time operator
  canonically conjugate to a Hamiltonian with non-empty point
  spectrum, Proc. Royal Soc. Lon. A {\bf 458}, 451 (2018).
\bibitem{hartle}J.B. Hartle, Quantum kinematics of spacetime. II. A
  model quantum cosmology with real clocks, Phys. Rev. D {\bf 38},
  2985 (1988).
\bibitem{damborena}J. A. Damborenea, I. L. Egusquiza, J. G. Muga, B.
  Navarro, Quantum dwell times, arXiv:quant-ph/0403081 (2004).
\bibitem{futuro}L. Maccone, K. Sacha, in preparation.
\bibitem{rovelliconst1}C.  Rovelli, Time in quantum gravity: An
  hypothesis, Phys. Rev.  D {\bf 43}, 442 (1991).
\bibitem{reirov}M. Reisenberger, C. Rovelli, Spacetime states and
covariant quantum theory, Phys. Rev. D {\bf 65}, 125016 (2002).
\bibitem{unruht}J. Oppenheim, B. Reznik, W. G. Unruh, When does a
  Measurement or Event Occur, Found. Phys. Lett. {\bf 13}, 107
  (2000), quant-ph/980064.
\bibitem{pauli} J. Leon, L. Maccone, The Pauli Objection, Found. Phys.
  {\bf 47}(12), 1597 (2017).
\bibitem{zurek}Robin Blume-Kohout, Wojciech H. Zurek, Quantum
  Darwinism: Entanglement, branches, and the emergent classicality of
  redundantly stored quantum information, Phys. Rev. A {\bf 73},
  062310 (2006).
\bibitem{giuli}D. Giulini, C. Kiefer, H.D.Zeh, Symmetries,
  superselection rules, and decoherence, Phys. Lett. A {\bf 199}, 291
  (1995). 
\bibitem{robertson}H.P. Robertson, The uncertainty principle, Phys.
  Rev. {\bf 34}, 163 (1929).
\bibitem{kuchar}K.V. Kucha\u r, ``Time and interpretations of quantum
  gravity'', Proc. 4th Canadian Conference on General Relativity and
  Relativistic Astrophysics, eds. G. Kunstatter, D.  Vincent, and J.
  Williams (World Scientific, Singapore, 1992), pg.~69-76.
\bibitem{vonneuman}J. von Neumann, {\em Mathematical Foundations of
    Quantum Mechanics} (Princeton Univ.  Press, 1955).
\bibitem{conttime}E. Moreva, G. Brida, M.  Gramegna, L. Maccone, M.
  Genovese, Quantum time: experimental multi-time correlations, Phys.
  Rev. D, {\bf 96}, 102005 (2017).
\bibitem{esperimentotorino} E. Moreva, G. Brida, M. Gramegna, V.
  Giovannetti, L. Maccone, M. Genovese, Time from quantum
  entanglement: an experimental illustration, Phys. Rev. A {\bf 89},
  052122 (2014).
\bibitem{wdw} B.S.  DeWitt, ``Quantum Theory of Gravity. I. The
  Canonical Theory'', Phys. Rev. {\bf 160}, 1113 (1967).
\end{references}
\end{document}